\documentstyle[12pt,epsfig]{article}
\newcommand{\be}{\begin{equation}}
\newcommand{\ee}{\end{equation}}
\newcommand{\ba}{\begin{eqnarray}}
\newcommand{\ea}{\end{eqnarray}}
\newcommand{\ci}[1]{\cite{#1}}
\newcommand{\lab}[1]{\label{#1}}
\begin{document}
%\draft

\title{ Features of high energy $pp$  \\
  elastic scattering   at small $t$}

  %\maketitle

  \author{ B. Nicolescu
  \\
 CNRS and Universit\'e Pierre et Marie Curie, Paris, \\
     e-mail: nicolesc@lpnhep.in2p3.fr \\
\phantom{} \\
  O.V. \ Selyugin   \\
 BLTP, JINR, Dubna, 141980 Russia  \\
    email: selugin@theor.jinr.ru }

  \date{}

%\author{B. Nicolescu$^{1}$ \thanks{nicolesc@lpnhep.in2p3.fr} ,
%  O.V. \ Selyugin$^{2}$  \thanks{selugin@thsun1.jinr.ru}}

%\affiliation{
%\address{ $^1$ \it Laboratoire de Physique Nucl\'eaire et des Hautes
%Energies (LPNHE) , \\
%CNRS and Universit\'e Pierre et Marie Curie, Paris }

%\author{ O. \ V. \ Selyugin$^{1}$ \thanks{selugin@thsun1.jinr.ru}}

%\address{$^2$ \it BLTP, JINR and Universit\'e  de Liege}

\maketitle

\abstract
 { A  method of  determination of the real part of the elastic scattering
  amplitude  is examined for high energy proton-proton
   elastic scattering at small momentum transfer. The method allows
   to decrease the number of model assumptions,
   to obtain the real parts  of the spin non-flip and spin-flip
 amplitudes in the narrow
   region  of momentum transfer. }

%\vspace{0.5cm}
\newpage

      A large number of experimental and theoretical studies of the
 high energy elastic
 proton-proton  and proton-antiproton scattering at  small angles
 gives a rich information about this  processes,
 which allows to narrow the circle of examined models
 and  to  solve a number of  difficult  problems.
 Especially this concerns  the  energy  dependence  of  some
 of characteristics of these reactions and the contribution of the odderon.

     Some  of these questions are connected with the $s$ and $t$ dependence
 of the spin-non-flip phase of hadron-hadron scattering.
 The majority  of the models define  the  real  part  of  the  scattering
 amplitude phenomenologically. Some  models  use  local
 dispersion relations       and   the   hypothesis   of   the   geometrical
 scaling. As it is known, using some  simplifying assumption, the
 information about   the   phase   of
 the   scattering   amplitude   can   be
 obtained from the experimental data at small momentum transfers   where
 the interference of the electromagnetic and hadron amplitudes takes  place.
 On the whole, the obtained information confirms the local dispersion
 relations.
  The knowledge of the structure of the elastic proton-nuclei and
  nuclear-nuclear scattering is needed to distinguish  different models
  describing high energy nuclei interactions. This is important
  for  the QCD approach of the high energy nuclei interaction
  \ci{armesto}.

     The standard procedure to extract the magnitude of the real part
  includes the fit of the experimental data taking the magnitude
   of total cross section, of the  slope, of $\rho$,
 and, sometimes of the normalization
  coefficient as free parameters.
\ba
 {\sum_{j}\chi_{j}^2} \ = \ \sum_{j=1}^{N} \frac{(
           \frac{
                  d\sigma}{dt}^{Exp.}_{j}  \ - \ \frac{d\sigma}{dt}^{Theor.}_{j})^2 }
 {  (\Delta^{Exp.}_{j})^2 }.
  \label{Rchi}
\ea

This procedure requires a sufficiently wide interval of $t$ and a large number
 of experimental points.

The spin-independent amplitude can be written
 as a sum of nuclear $\Phi^h(s,t)$
and electromagnetic  $\Phi^e(s,t)$ amplitudes :
\ba
\Phi(s,t)=\Phi^h(s,t)+e^{i\alpha\varphi}\Phi^e(s,t)\ .
\ea
where $\Phi^e(s,t)$ are the leading-terms at high energies of the one-photon
amplitudes as defined, for example,
 in \ci{leader} and the common phase $\varphi$     is
\ba
\varphi=-[\gamma+\log\big(B(s,t)\vert t\vert/2\big)+\nu_1+\nu_2]
\ea
where $B(s,t)$ is the slope of the nuclear amplitude,
 $\gamma = 0.577$,
 and $\nu_1$ and $\nu_2$ are
small correcting terms  define the behavior of the Coulomb-hadron phase
at small momentum transfers (see \ci{selprd}).
At very small t and fixed s, these electromagnetic amplitudes are such that
$
\Phi_1^e(s,t) = \Phi_3^e(s,t)\sim\alpha/t \ ,
\Phi_2^e(s,t) = -\Phi_4^e(s,t)\sim \alpha\cdot \hbox{const.}\ ,
\Phi_5^e(s,t)  \sim  -\alpha/\sqrt{\vert t\vert}\ .
$, where $\alpha$ is the fine-structure constant.
  We assume, as usual, that at high energies and small angles
  the double-flip amplitudes are small with respect to the spin-nonflip one
  and that spin-nonflip amplitudes are approximately equal. Consequently,
  the observables  are determined by two amplitudes:
  $ F (s,t) = \Phi_{1}(s,t) + \Phi_{3}(s,t)
   =F_{N}+F_{C} exp(\ {rm i} \alpha \varphi)$.

    In the standard fitting procedure
\ba
d\sigma/dt &= \pi [ (F_{C} (t))^2
%          + (Re F_{N}(s,t))^{2}+ (Im F_{N}(s,t))^{2})      \nonumber \\
          + (\rho(s,t)^2 + 1) (Im F_{N}(s,t))^{2})      \nonumber \\
 &+ 2 (\rho(s,t)+ \alpha \varphi(t)) F_{C}(t) Im F_{N}(s,t)] . \label{ds2}
\ea
 $F_{C}(t) = \mp 2 \alpha G^{2}/|t|$ is the Coulomb amplitude;
  and $G^{2}(t)$ is  the  proton
electromagnetic form factor squared.
$Re\ F_{N}(s,t)$ and $ Im\ F_{N}(s,t)$ are the real and
imaginary parts of the nuclear amplitude and
$\rho(s,t) = Re \ F_{N}(s,t) / Im \ F_{N}(s,t)$.
 The formula (3) is used for the fit  of  experimental  data
determining the Coulomb and hadron amplitudes and the Coulomb-hadron
phase in order  to obtain the value of $\rho(s,t)$.

        $Re F_{N}(s,t)$ is obtained by fitting
  the differential cross sections either taking into account
 the value of $\sigma_{tot}$ from another experiment,  as made
 by the UA4/2 Collaboration, or taking  $\sigma_{tot}$ as a free
 parameter, as made in \ci{selpl}.
 If one does not take the normalization coefficient as  a free parameter in
 the fitting procedure, its definition requires the knowledge of
 the behavior of imaginary and real parts of the scattering amplitude
 in the range of small transfer momenta and the magnitude of
 $\sigma_{tot}(s)$ and $\rho(s,t)$.

     In this talk, we briefly describe  some new procedures of simplifying
 the determination of elastic scattering amplitude parameters.

   From equation   (\ref{ds2})
    one can obtain the equation for  $Re F_{N}(s,t)$
   for every experimental point $i$
\ba
  && Re F_{N}(s,t_i)= -Re F_{C}(s,t_i)   \nonumber    \\
  & & \pm [(1+\delta) / \pi d\sigma/ dt(t=t_i)
   - (\alpha \phi F_{C}(t_i)+Im F_{N}(t_i))^2]^{1/2}.
                                               \label{rsq}
\ea
 where $\delta$ is the corrections coefficient which reflect the accuracity
  of the normalization parameter $n=1+\delta$.
% our determination of $d\sigma /dt^{exp}$.
 As the imaginary part of scattering amplitude is defined by
\be
  Im F_{N}(s,t) = \sigma_{tot}/(0.389 \cdot 4 \pi) exp(B/2 t), \lab{im}
\ee
 it is obvious from  (\ref{rsq}) that the determination of
 the real part depends on
 $n, \sigma_{tot}, B$.
     The magnitude of $\sigma_{tot}$ determined from experimental
   data depends on the normalization parameter $n=1+\delta$
  which reflects
   the experimental error in determining $d\sigma/dt$ from $dN/dt$.
      The equation (\ref{rsq}) shows  the possibility to calculate
  the real part in every separate point of $t_i$ if the imaginary part of
   scattering amplitude and $n$ are determined and to check up the form
  of the obtained  real part of the scattering amplitude or vice versa
  (see \cite{selyf}).
  This form shows also a minimum value of $n$, as the expression
  situated under the square root cannot be less then zero.

   Let us define the sum of the real part hadron and Coulomb amplitudes
  as $\Delta_{R}$, so we can write:
\ba
  && \Delta_{R}(t_i) =  [ Re F_{N}(s,t_i)+Re F_{C}(s,t)]^2 =  \nonumber    \\
  & &  [(1+\delta)/\pi \  d\sigma / dt(t=t_i)
   - (\alpha \phi F_{C}(t_i)+Im F_{N}(t_i))^2].
                                               \label{Del}
\ea
        This formula shows a significant property for
  the proton-proton cross section at a very high energy and proton-antiproton
  scattering at low energy, where the real part
  of the hadronic amplitude is sufficiently large and is opposite in sign
  relative to the Coulombic amplitude.
  We therefore get
 \ba
  && \Delta_{R}(t_i) =  (1+\delta)( Re F_{N}(s,t_i)+Re F_{C}(s,t))^2   \nonumber  \\
  & &
   - \delta (\alpha \phi F_{C}(t_i)+Im F_{N}(t_i))^2].
                                               \label{Del1}
\ea

      Let us examine this expressions for the $pp$-scattering at energies
      above $\sqrt{s} = 540 \ GeV$ where the real part of the hadron
      spin-non-flip amplitude is positive and non-negligible.
    For this aim, let us make a gedanken experiment and calculate
  $d\sigma/dt$
  with definite parameters ($\rho= 0.15$ and $\sigma_{tot}=63$
  taking them as  experimental points.
% of the differential cross sections.
%  In this case, we know exactly what we obtain at the end of our
%  calculation.
  For the $pp$-scattering at high energies,
  the equation (\ref{rsq})
% , as the old form (\ref{ds0}),
  has a remarkable property.

%%%%%% FIG.1

  The real part of the Coulomb scattering amplitude of $pp$-scattering
  is negative and
  exceeds the size of $F_h(s,t)$ at
  $t \rightarrow 0$ , but has a large slope.
   As the real part of the hadronic amplitude is positive at high energies,
%   so, it is obviously,
  it results that
% that this expression  -
$\Delta_R$ has a minimum situated
  in  $t$  independent of $n$ and $\sigma_{tot}$ as shown
 in Fig. 1.

  The position of the minimum  gives us the value
  $t_{R}$ where $Re F_{N} = -Re F_{C}$. As we know the
 Coulomb amplitude, we estimate the real part of the
 $pp$-scattering amplitude at this point. Note that all other methods give us
 the real part only in a sufficiently wide interval of the transfer
 momenta.
    If we choose  the right normalization coefficient and $\sigma_{tot}$
  our minimum will be equal zero. But if the normalization coefficient
   is not right one the minimum will be or above or lower then zero,
  but practically it is located
  at the same point $t_R$. So, the size of $\Delta_R$ shows us  the
  valid determination of the normalization coefficient and $\sigma_{tot}$.

  This method works only in the case of the positive real part
 of the nucleon amplitude, and it is especially
  good in the case of large $\rho$. So, it is
 interesting for the  experiment $PP2PP$ at RHIC and the future
  TOTEM experiment   at LHC.

    Though in the range of ISR we have small $\rho(s, t \approx 0)$ and
 few experimental points, let us try to examine one experiment,
 for example, at $\sqrt{s}=52.8\ GeV$. This analysis is shown in Fig. 2.
 One can see that in this case the minimum is sufficiently large, and
 $-t_{min}= (3.3 \pm 0.1)10^{-2} \ GeV^2$.
 The corresponding real part
 equals $0.442 \pm 0.014 \ GeV$.

% If we take, as in the experiment,
 Our analysis gives
 $\rho=0.063 \pm 0.003$, while   the paper \ci{528} gives
 $\rho=0.077 \pm 0.08$.

%%%%%%%FIG.2.
%\vspace{-1cm}

  For RHIC energies we can simulate the "experimental" data taking
  the calculated theoretical curve with  certain parameters
  $\sigma_{tot}, B, \rho$ and the magnitude of errors which
  are expected in the future experiment. Then we calculate the deviation
  from the theoretical curve in  units of errors using a Gaussian random
  procedure in order
  to calculate  the probability of the deviation by a number
  of errors.
  As a result, we obtain
  the differential cross sections modeling the future experimental data,
  for example,  with the posible size of
  $\rho = 0.135$ and  $\rho=0.175$.
  Then we can  determing the value of $\Delta_R$ from
   these  ``gedanken'' experimental data , which are shown in Fig.3 (a,b)
   correspondingly.
   The difference between these two
    modeling data representations is obvious.
   The pure theoretical representation of $\Delta_R$
   with the same values of $\rho$ and
    with $\rho=0$ are shown also.

      Our predictions for the LHC energies are shown in Fig.4 and 5. for the
      value of $\rho=0.15$.

   Note that the point  $t_{R}$ is
  important for
   the determination of the real part of spin-flip amplitude also
  \cite{PS1}.
 At high energies and small angles the analyzing power
can written  in form
\ba
 - A_N\frac{d\sigma}{dt}/2& =&
   ImF_{nf}^h (ReF_{sf}^c +  ReF_{sf}^{h}) +
Im F^{c}_{nf} (Re F^{c}_{sf}  + Re F^{h}_{sf})  \\ \nonumber
   &&  - ImF_{sf}^c ( Re F_{nf}^{c} + Re F_{nf}^{h})
     - ImF_{sf}^h (ReF_{nf}^c  + ReF_{nf}^h ).    \\ \nonumber
\ea

 We obtain for proton-proton scattering at high energies at the point  $t_{R}$
 where $Re F^{nf}_h = - Re F^{nf}_c$,
\ba
  Re F_{sf}^{h}(s,t) = \frac{-1}{2  (Im F_{nf}^{h}(s,t)+Im F_{nf}^{c}(t))}
  A_{N}(s,t) \
    \frac{d\sigma}{dt} \
	       - Re F_{sf}^{c}(t).   \lab{refhm}
\ea
  At this point some terms in the definition of analyzing power
  will be canceled. Such a representation can be used
  for the determination of the real part of the hadron spin-flip
  amplitude at high energy and small angles.

   It is interesting to apply this new method to the proton-nuclear
   scattering at high energies.
    The size of the hadron amplitude grows only slightly
   less then proportional to $A$. If $\sigma_{tot}(pp) = 38 \ $ mb
   in the region of hundred GeV, the  $\sigma_{tot}(p ^{12}C) = 335 \ $mb.
   The most important difference with
  $pp$-scattering is that the slope is very high, near $70 \ mb/GeV^2$.
%   of this nuclear reaction at hundred GeVs.
  The electromagnetic amplitude grows like $Z$
  and its slope also grows. It is interesting that the simple calculations
  which take the hadron amplitude  at small momentum transfer
  in the usual exponential form with
  large slope leads  the practically the same results as for the
  proton-proton scattering.

 The precise experimental measurements of $dN/dt$ and $A_N$ at
 RHIC, as well as, if possible, at the Tevatron, will therefore give us
 unavailable information on the hadron elastic scattering at small t. New
 phenomena at high energies could be therefore detected without going
 through the
 usual arbitrary assumptions (such as the exponential form)
 concerning the hadron elastic scattering amplitudes.
  It is  interesting to apply this method to the proton-nuclei
  scattering at high energies, especially at RHIC energies. This
  method offers
  a unique possibility search for  the behavior of the real part
  of the hadronic amplitude in  nuclear reactions.

\newpage

%     {\it Acknowledgment.} {\hspace {0.5cm} The authors express his deep
%gratitude to J.-R. Cudell, J. Cugnon, W. Guryn and E. Martynov
%for fruitful discussions. One of us (O. S.) thanks Prof. Jean-Eudes Augustin
%for the hospitality at the LPNHE Paris, where part of this work was done.

\newpage
     Figure  captions

FIG.1.
 The model calculations of  $\Delta_R$
 for the  $pp$-scattering
 at RHIC energy $\sqrt{s}=540 \ GeV$  and with $\sigma_{tot} = 63 \ mb$
 and  different $n$.
%  and  $\sigma_{tot}$ Eq.(\ref{im})\\
% a) with $\sigma_{tot}=63 \ mb$; b) with $\sigma_{tot}=62 \ mb$.

FIG.2.
  The calculation of $\Delta_R$ for the
$pp$-scattering
 using the experimental data of $d\sigma/dt$ at $\sqrt{s}=52.8 \ GeV$
  \cite{528}. The lines are the polynomial fit of the points calculated
  with experimental data and with different $n$.

FIG.3.
  The  calculation of $\Delta_R$ for the $pp$-scattering
 at RHIC  with a) $\rho_1 =0.135$ and b) $\rho_2 =0.175$
 The solid, short-dashed, and dotted lines are the
 theoretical curves for $\rho_2 = 0.175$, $\rho_1 = 0.135$,
   and  $\rho_0 = 0$   respectively.

FIG.4.
  The  calculation of $\Delta_R$ for the $p ^{12}C$ -scattering
   with  $\rho =0.1$ and  $\rho =0.075$
 (the solid and  dashed lines, correspondingly)
  using  of the exponential behavior of the hadronic amplitude.

FIG.5.
  The  calculation of $\Delta_R$ for the $pp$ -scattering
   with  $\rho =0.15$ at $\sqrt{s}= 4 \ $TeV

FIG.6
 The  calculation of $\Delta_R$ for the $pp$ -scattering
   with  $\rho =0.15$ at $\sqrt{s}= 14 \ $TeV

\newpage

%\label{fig1}
\begin{figure}[!ht]
%\vskip -1.5cm
\epsfysize=90mm
%\centerline{\epsfbox{mpp54063.ps}}
\centerline{\epsfbox{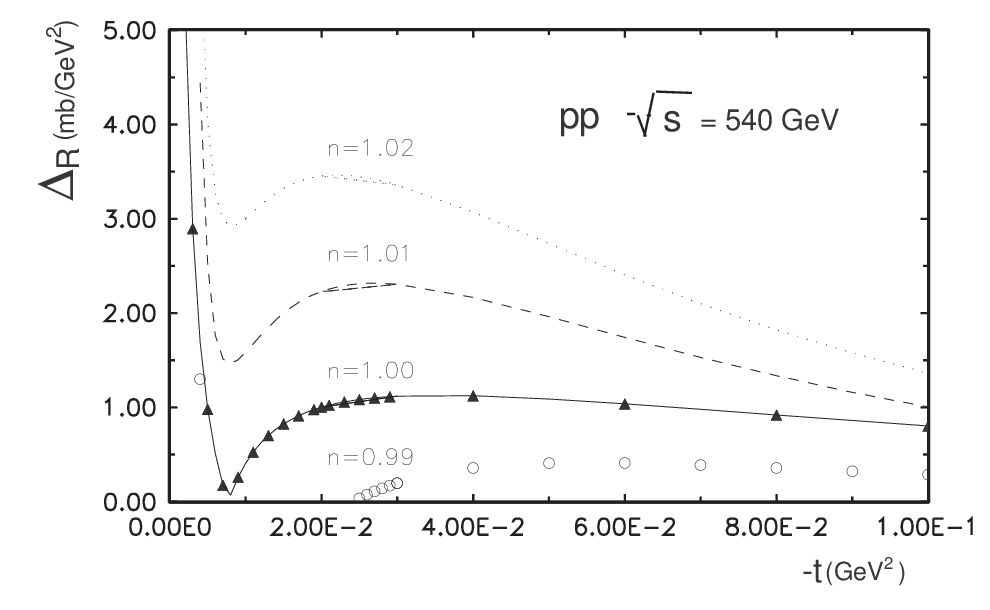}}
\vspace{-0.5cm}
\centerline{Fig.1 }
%\vskip -1.cm
\epsfysize=80mm
\centerline{\epsfbox{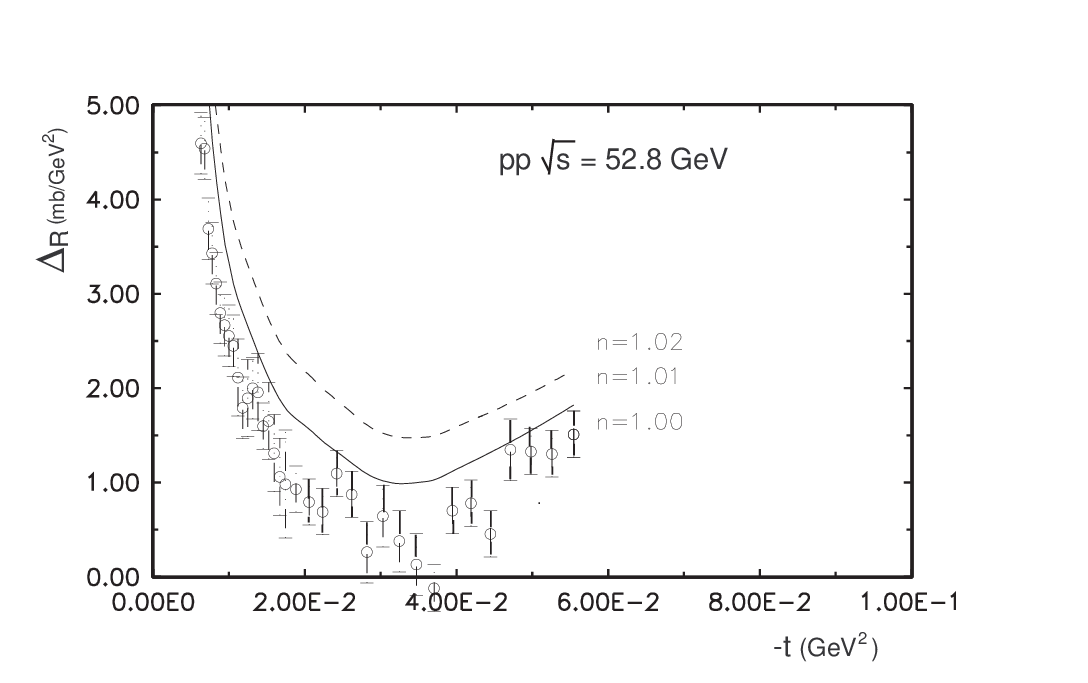}}
\vspace{-0.5cm}
\centerline{Fig. 2}
%\caption{
% The model calculations of the $del$
% for the
% $pp$-scattering
% at RHIC energy $\sqrt{s}=540 GeV$ on different $n$ \\
% a) with $\sigma_{tot}=63 \ mb$; b) with $\sigma_{tot}=62 \ mb$.
%}
\end{figure}

\newpage

%\label{fig3}
\begin{figure}[!ht]
\epsfysize=80mm
\centerline{\epsfbox{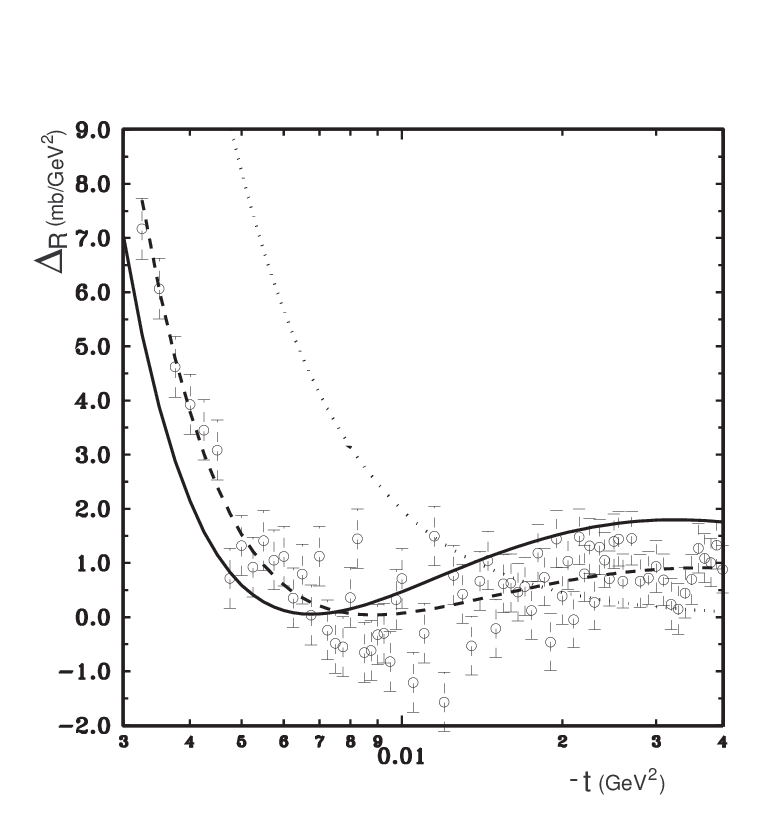}}
\vspace{-0.5cm}
\centerline{Fig.3 a}
\epsfysize=80mm
\centerline{\epsfbox{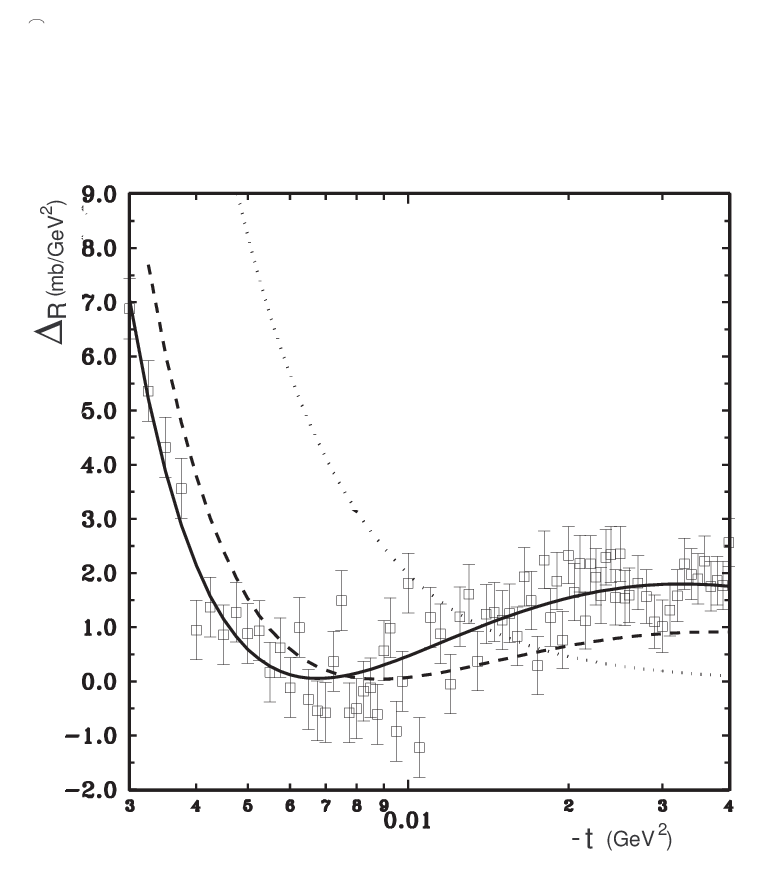}}
\vspace{-0.5cm}
\centerline{Fig.3 b}
%\vspace{-0.5cm}
%\caption{
%  The  calculation of $del$ for the model $pp$-scattering
% with a) $\rho_1 =0.135$ and b) $\rho_2 =0.175$
% The solid, short-dashed, and dotted lines are the
% theoretical curves for $rho_2 = 0.175$, $rho_1 = 0.135$, $rho_0 = 0.$
%  respectively.
%}
\end{figure}
%\vskip -0.5cm

\newpage

%\label{fig3}
\begin{figure}[!ht]
\epsfysize=80mm
\centerline{\epsfbox{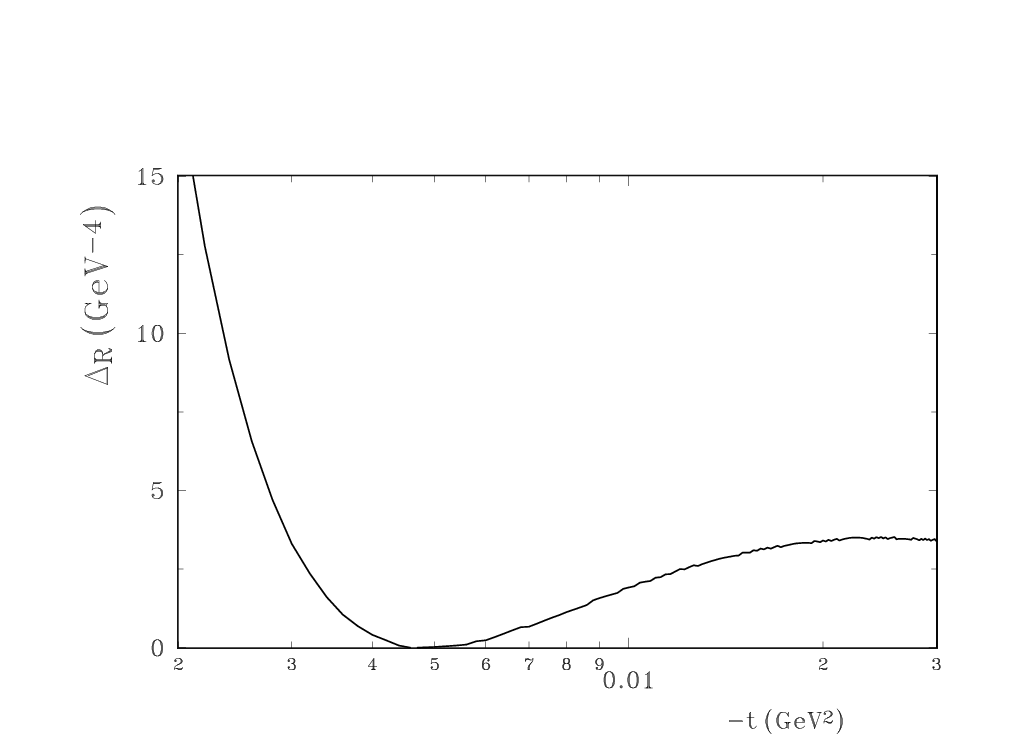}}
\vspace{-0.5cm}
\centerline{Fig.4 }
\epsfysize=80mm
\centerline{\epsfbox{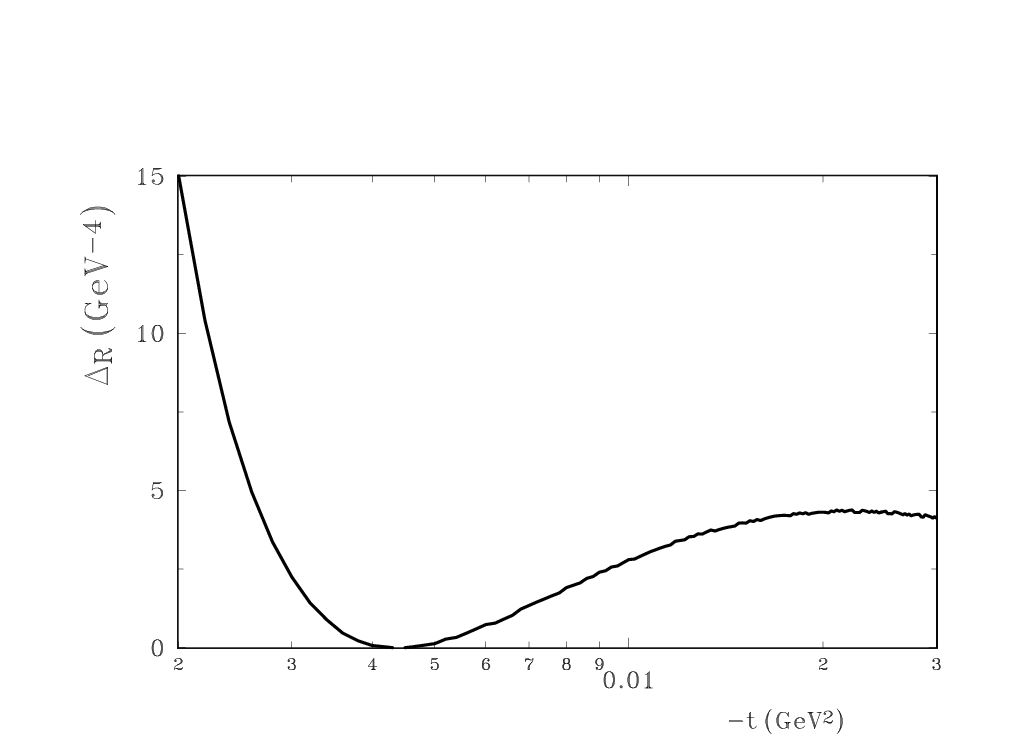}}
\vspace{-0.5cm}
\centerline{Fig.5}

\end{figure}
%\vskip -0.5cm

\end{document}